# Multiscale Thermodynamics: Energy, Entropy, and Symmetry from Atoms to Bulk Behavior

Ralph V. Chamberlin [1,*], Michael R. Clark [1], Vladimiro Mujica [2] and George H. Wolf [2]

1. Department of Physics, Arizona State University, Tempe, AZ 85287-1504 USA; chamberl@asu.edu, mrclark5@asu.edu
2. School of Molecular Science, Arizona State University, Tempe, AZ 85287-1604 USA; vmujica@asu.edu, gwolf@asu.edu
* Correspondence: chamberl@asu.edu

**Abstract:** Here we investigate how the local properties of particles in a thermal bath may influence the thermodynamics of the bath, and consequently alter the statistical mechanics of subsystems that comprise the bath. We are guided by the theory of small-system thermodynamics, which is based on two primary postulates: that small systems can be treated self-consistently by coupling them to an ensemble of similarly small systems, and that a large ensemble of small systems forms its own thermodynamic bath. We adapt this "nanothermodynamics" to investigate how a large system may subdivide into an ensemble of smaller subsystems, causing internal heterogeneity across multiple size scales. For the semi-classical ideal gas, maximum entropy favors subdividing a large system of "atoms" into an ensemble of "regions" of variable size. The mechanism of region formation could come from quantum exchange symmetry that makes atoms in each region indistinguishable, while decoherence between regions allows atoms in separate regions to be distinguishable by their distinct locations. Combining regions reduces the total entropy, as expected when distinguishable particles become indistinguishable, and as required by a theorem in quantum mechanics for sub-additive entropy. Combining large volumes of small regions gives the usual entropy of mixing for a semi-classical ideal gas, resolving Gibbs paradox without invoking quantum symmetry for atoms that may be meters apart. Other models presented here are based on Ising-like spins, which are solved analytically in one dimension. Focusing on the bonds between the Ising-like spins we find similarity in the equilibrium properties of a two-state model in the nanocanonical ensemble and a three-state model in the canonical ensemble. Thus, emergent phenomena may alter the thermal behavior of microscopic models, and the correct ensemble is necessary for fully-accurate predictions. Another result using Ising-like spins involves simulations that include a nonlinear correction to Boltzmann's factor, which mimics the statistics of indistinguishable states by imitating the dynamics of spin exchange on intermediate lengths. These simulations exhibit $1/f$-like noise at low frequencies ($f$), and white noise at higher $f$, similar to equilibrium thermal fluctuations found in many materials.

**Keywords:** nanothermodynamics; fluctuations; maximum entropy; finite thermal baths; corrections to Boltzmann's factor; ideal gas; Ising model; Gibbs' paradox; statistics of indistinguishable particles

## 1. Introduction

Thermodynamics and statistical mechanics provide two theoretical approaches for interpreting the thermal behavior shown by nature [1–7]. In statistical mechanics, the local symmetry of particles in the system is well known to influence their behavior, yielding Maxwell-Boltzmann, Bose-Einstein, or Fermi-Dirac statistics. Lesser known is the impact of local symmetry on thermodynamics. Difficulty in understanding local behavior in standard thermodynamics is due to its two basic postulates: that all systems must be macroscopic, and homogeneous. Hence, accurate theories of local thermal properties require that finite-size effects are added to thermodynamics, such as in fluctuation theorems [8,9] and stochastic thermodynamics [10,11]. These approaches are remarkably successful at describing thermal dynamics, including far-from-equilibrium processes. However, in general these theories require that, at least at some time during the dynamics, the system must have ideal thermal contact to an ideal heat bath; specifically, the system must be weakly, rapidly, and homogeneously coupled to an effectively infinite heat reservoir. Here we focus on the theory of small-system thermodynamics [12–14], which is based on two novel postulates: that small systems can be treated self-consistently by coupling them to similarly small systems (without correlations between the small systems so that they form an ensemble), and that a large ensemble of small systems becomes its own effectively infinite heat reservoir. Thus, this "nanothermodynamics" provides a fundamental foundation for connecting thermal properties across multiple size scales: from microscopic particles, through mesoscopic subsystems, to macroscopic behavior.

Although small-system thermodynamics was originally developed to treat individual molecules and isolated clusters, we use nanothermodynamics as a guide to study how large systems are influenced by internal heterogeneity, especially on the scale of nanometers [15–17]. Here we focus on how thermal equilibrium often involves subdividing a large system into an ensemble of subsystems, usually requiring that the number of particles and volume of each subsystem fluctuate freely, without external constraint. We call these freely-fluctuating subsystems "regions," and the self-consistent system of independently-fluctuating nanometer-sized regions the "nanocanonical" ensemble.

A common assumption in standard thermodynamics is additivity. For example, it is assumed that if the size of a system is doubled, then all of its extensive variables (e.g. internal energy $E$, entropy $S$, and number of particles $N$) will precisely double, and all of its intensive variables (e.g. temperature $T$, pressure $P$, and chemical potential $\mu$) will remain unchanged. In fact, the Gibbs-Duhem relation (found from assuming that all extensive variables increase linearly with $N$) is often used as a test of consistency in thermodynamics. Thus, Gibbs' paradox [18–21] comes from the apparent discrepancy between the predictions of thermodynamics and classical statistical mechanics. Specifically, standard thermodynamics states that when a partition is reversibly removed between two identical systems the total entropy should be exactly twice the entropy of each initial system, while Maxwell-Boltzmann statistics of a classical ideal gas predicts an additional term proportional to $N\ln(2)$ due to the entropy of mixing. Here we obtain several results where various thermodynamic quantities (including $E$, $S$, and $\mu$) contain terms that depend nonlinearly on $N$, and we focus on how these finite-size effects are necessary to conserve total energy and maximize total entropy.

The broad generality of thermodynamics can be a distraction to many scientists who prefer the concrete models and microscopic details that constitute statistical mechanics. For similar reasons, statistical mechanics is often said to be a foundation for thermodynamics. However, the fundamental physical laws are in the thermodynamics, and strict adherence to these laws is necessary before statistical mechanics can fully represent the real world. Here we briefly explain how the laws of thermodynamics can be extended to length scales of nanometers, and why applying the resulting nanothermodynamics to statistical models may improve their accuracy and relevance to real systems.

## 2. Background

*2.1 Standard thermodynamics*

Thermodynamics establishes equations and inequalities between thermodynamic variables which must be obeyed by nature. Thermodynamic variables usually come in conjugate pairs, whose product yields a contribution to the internal energy of a system. The fundamental equation for reversible changes in thermodynamics (aka the Gibbs equation, which combines the first and second laws) states that the total internal energy of the system can be changed by changing one (or more) of the thermodynamic variables. For example, the fundamental equation of $N$ particles in volume $V$ is

$$dE = TdS - PdV + \mu dN. \tag{1}$$

Equation (1) gives three ways to reversibly change the internal energy of a system ($dE$): heat can be transferred ($TdS$), work can be done ($PdV$), or the number of particles can be changed ($\mu dN$).

The primary conjugate variables in Eq. (1) used to define thermal equilibrium [1] are $T$ and $S$. Temperature is the familiar quantity that is "hotness measured on some definite scale" [4], but entropy is a more-subtle concept [22–25]. Although entropy is often associated with randomness, any system with sufficient information about the randomness may also have low entropy. Thus, a more-general definition of entropy comes from quantifying the amount of *missing* information. Entropy is a main focus of our current study, especially in how total entropy can be maximized by proper choice of thermodynamic ensemble.

Equation (1) describes changes to a fundamental thermodynamic function ($E$), with $dE = 0$ when internal energy is conserved, characteristic of the microcanonical ensemble that applies to isolated systems. Fundamental functions in other standard ensembles are free energies that also do not change in the relevant equilibrium. The ensemble that is most appropriate to a system depends on how it is coupled to its environment. In general, each pair of conjugate variables that contribute to the internal energy in Eq. (1) includes an "environmental" variable (controlled by the environment, e.g. researcher) and its conjugate that responds to this control. Different sets of environmental variables form distinct thermodynamic ensembles. Different ensembles may be connected to statistical mechanics using ensemble averages that are equated to thermodynamic quantities. In principle, for simple systems having 3 pairs of conjugate variables there are eight ($2^3$) possible ensembles [26]. In practice, however, only seven of these ensembles are well-defined in standard thermodynamics. Examples include the fully-closed microcanonical ensemble, as well as the partially-open ensembles: canonical, Gibbs', and grand-canonical. The fully-open generalized ensemble, involving three intensive environmental variables (e.g. $\mu$, $P$, $T$) is ill-defined in standard thermodynamics because at least one extensive environmental variable is needed to control the size of the system. Thus, nanothermodynamics is the only way to treat fully-open systems in a consistent manner, allowing the system to find its equilibrium distribution of subsystems without external constraints. In fact, the nanocanonical ensemble is necessary for the true thermal equilibrium of any system having independent internal regions, especially from localized internal fluctuations. Because large and homogeneous systems all yield equivalent behavior for all ensembles, it is sometimes said that the choice of ensemble can be made merely for convenience [14]. However, for small systems, and for bulk samples that subdivide into an ensemble of small regions, the choice of ensemble is crucial, so that the correct ensemble must be used for realistic behavior. Indeed, the correct ensemble is essential for fully-accurate descriptions of fluctuations, dynamics, and the distribution of independent regions inside most samples [15–17].

*2.2 Standard statistical mechanics*

The usual foundation for statistical mechanics is Boltzmann's factor, $e^{-E/kT}$. This $e^{-E/kT}$ is used to obtain the probability of finding states of energy $E$ at temperature $T$, yielding Maxwell-Boltzmann statistics for distinguishable particles, Bose-Einstein statistics for symmetrical bosons, and Fermi-Dirac statistics for anti-symmetrical fermions. This $e^{-E/kT}$ also provides the foundation for most modern results involving stochastic thermodynamics and fluctuation theorems [8-11,27–32]. However, $e^{-E/kT}$ is based on several assumptions [33–35]. Specifically, the degrees of freedom must be in ideal thermal contact with an ideal heat bath; i.e. there must be fast (but weak) thermal contact to a homogeneous and effectively infinite heat reservoir. Whereas several experimental techniques have shown that most primary degrees of freedom couple slowly to the heat reservoir, with energies that are persistently localized on time-scales of the primary response (e.g. 100 s to 100 μs) [36–42]. Furthermore, other techniques [43–48] have established that this localization involves regions with dimensions on the order of nanometers (e.g. 10 molecules to 390 monomer units [49]). Thus, the basic requirements of standard statistical mechanics for fast coupling to a homogenous heat reservoir are absent for the primary response in most materials, including liquids, glasses, polymers, and crystals. Moreover, molecular dynamics (MD) simulations of crystals with realistic interactions exhibit excess energy fluctuations that diverge like $1/T$ as $T \rightarrow 0$, deviating from standard statistical mechanics based on $e^{-E/kT}$, but quantitively consistent with energy localization on length scales of nanometers [50]. Nanothermodynamics is necessary to treat independent subsystems that have a distribution of sizes, with realistic particles and heterogeneous interactions, allowing the laws of thermodynamics that govern statistical mechanics to be extended across multiple size scales, down to individual atoms.

Entropy in statistical mechanics usually involves calculating the number of distinct microscopic states that yield the same macroscopic state. This may be expressed in terms of the multiplicity ($\Omega_i$) of microstates that yield the $i^{th}$ macrostate, or in terms of the average probability of finding the $i^{th}$ macrostate ($\rho_i$). The Gibbs (or Boltzmann-Gibbs) expression for average entropy is given by $\bar{S}/k = -\sum_i \rho_i \ln \rho_i$, where $k$ is Boltzmann's constant and the sum is over all possible macrostates. Alternatively, Boltzmann's expression for the entropy of each macrostate is $S_i/k = \ln(\Omega_i)$. In the microcanonical ensemble, where all microstates are assumed to be equally likely, both expressions yield identical values for equilibrium average behavior $\bar{S} = \sum_i S_i$. (Generalized entropies have been introduced to investigate the possibility that all microstates are not equally likely [51], but here we focus on specific non-additive and non-extensive contributions to entropy that arise naturally from finite-size effects in thermodynamics.) Gibbs' expression has the advantage that it also applies to other macroscopic ensembles, while Boltzmann's expression has the advantage that it can accommodate non-equilibrium conditions [52], including small systems that may fluctuate. Here we stress how nanoscale thermal properties must also govern large systems that subdivide into a heterogeneous distribution of subsystems, and how nanothermodynamics impacts the statistical mechanics of specific models.

The remainder of this overview is organized as follows. Section **3** is an introduction to nanothermodynamics. Section **4** is a review of how standard statistical mechanics is extended by nanothermodynamics. In section **5**, we apply nanothermodynamics to *5.1* the semi-classical ideal gas, and *5.2–5.6* various forms of Ising-like models for binary degrees of freedom ("spins") in a one-dimensional (1-D) lattice. In section **6** we conclude with a brief summary.

## 3. An Introduction to Nanothermodynamics

The primary postulate in standard thermodynamics that systems must be homogeneous fails to account for measurements on most types of materials that show thermal and dynamic heterogeneity [36–49]. We argue that this inability to explain heterogeneity can be traced to sources of energy and entropy that arise from finite-size effects, especially sources that occur on the scale of nanometers, which require nanothermodynamics to be treated in a self-consistent and complete manner.

After Gibbs introduced the chemical potential in 1876 it was widely believed that all categories of thermal energy were included in Eq. (1); at least for a simple gas. However, in 1962 Hill made a similar modification to the fundamental equation of thermodynamics when he introduced the subdivision potential, $\mathcal{E}$, and the number of subsystems, $\eta$. One way to understand $\mathcal{E}$ is to compare it to $\mu$. The chemical potential is the change in (free) energy to take a single particle from a bath of particles into the system, whereas $\mathcal{E}$ is the additional change in energy to take a cluster of interacting particles from a bath of clusters into the system, and in general $N$ interacting particles do not have the same energy as $N$ isolated particles, due to surface effects, length-scale terms, finite-size fluctuations, local symmetry, etc. Many finite-size contributions to energy can be included in the net Hamiltonian of the system. Indeed, already in 1872 Gibbs included surface energies proportional to $N^{2/3}$, and recent results have greatly extended these ideas to include shape-dependent terms [53–55]. Here we focus on contributions to $\mathcal{E}$ that are not explicitly contained in the microscopic interactions, instead emerging from nanoscale behavior. One example is the semi-classical ideal gas, where $\mathcal{E}$ comes entirely from entropy due to indistinguishable statistics of particles within regions, with particles in separate regions distinguishable by their locations. Other examples utilize 1-D systems, where the increase in entropy favoring a distribution of localized fluctuations can dominate the decrease in energy favoring uniform interactions, so that thermal equilibrium often involves dynamic heterogeneity. Because $\mathcal{E}$ uniquely accommodates contributions to energy and entropy from local symmetry and internal fluctuations, $\mathcal{E}$ is essential for strict adherence to the laws of thermodynamics across all size scales, especially behavior that emerges on the scale of nanometers.

Adding Hill's pair of conjugate variables to Eq. (1), using the formal definition of the subdivision potential $\mathcal{E} = (\partial E_t / \partial \eta)_{S_t, V_t, N_t}$, the fundamental equation of nanothermodynamics becomes [12–14]

$$dE_t = TdS_t - PdV_t + \mu dN_t + \mathcal{E}d\eta. \qquad (2)$$

Here, subscript $t$ denotes extensive variables for the large ensemble of small systems, which are related to the corresponding quantities for small systems ($E, S, V,$ and $N$) by $E_t = \eta E$, $S_t = \eta S$, $V_t = \eta V$, and $N_t = \eta N$. In Eq. (2), $\mathcal{E} d\eta$ adds finite-size effects as a way that internal energy can be changed, even for large systems. As examples, when a large system is subdivided into smaller subsystems, energy can change due to interface energies, excitation confinement, increased fluctuations, local symmetry, etc. More importantly, a large system can subdivide itself into an equilibrium distribution of nanoscale regions (the nanocanonical ensemble), which yields heterogeneity consistent with many measurements [36–49]. Although finite-size effects in internal energy cause non-intensive $\mu$ and contributions to $S$ that depend nonlinearly on $N$, the subdivision potential also includes emergent finite-size effects that are unique to nanothermodynamics. One way to obtain $\mathcal{E}$ is to integrate Eq. (2) from $\eta = 0$ to $\eta$ (with $T$, $P$, and $\mu$ held constant), then divide by $\eta$, yielding the Euler equation for nanothermodynamics [12–14]

$$E = TS - PV + \mu N + \mathcal{E}. \tag{3}$$

Thus, although $\mathcal{E}$ is never an extensive variable, it is added to extensive terms to give the total internal energy. Combining Eqs. (2) and (3) yields the Gibbs-Duhem equation for nanothermodynamics $d\mathcal{E} = -TdS + VdP - Nd\mu$. Both $\mathcal{E}$ and $d\mathcal{E}$ are always negligible in homogeneous and macroscopic systems. However, heterogeneity and other finite-size effects usually involve non-extensive contributions to energy and entropy, so that nonzero $\mathcal{E}$ and $d\mathcal{E}$ must be considered for fully-accurate results.

In ideal gases and other non-interacting systems, $\mathcal{E}$ comes entirely from entropy. For interacting systems, subdividing into smaller regions adds interfaces that often increase the energy; but the increase in energy from interfaces can be relatively small for large regions, so that energy reductions from finite-size effects and increased entropy from added configurations can favor subdivision. Examples of energy reductions include added surface states and increased fluctuations towards the ground state. In fact, $\mathcal{E}$ provides the only systematic way to maximize total entropy and consider all contributions to energy that emerge from nanoscale thermal fluctuations, surface states, and normal modes that are influenced by transient disorder. Furthermore, $\mathcal{E}$ facilitates treating local symmetry by using the statistics of indistinguishable particles within each region, while particles in neighboring regions are distinguishable by their separate locations. Solving Eq. (3) for the entropy yields

$$S/k = (E + PV - \mu N - \mathcal{E})/kT. \tag{4}$$

In much of Hill's work, he focused on average contributions to $\mathcal{E}$ from small-number statistics and surface effects in small systems, which can also be treated by including appropriate terms in the Hamiltonian. Here we focus on how his small-system thermodynamics can be adapted to treat subsystems from nanoscale heterogeneity inside bulk samples. In general, this nanothermodynamics may include contributions to energy that do not appear explicitly in the Hamiltonian, instead emerging from nanoscale behavior in a way that requires nanothermodynamics for a full treatment.

Figure 1 depicts four models for how a system can be subdivided into $\eta = 9$ subsystems [17]. Each model assumes subsystems that are small (e.g. on the scale of nanometers), and uncorrelated with other subsystems to yield an ensemble of small systems, which differs from ensembles of large systems often used to develop standard thermodynamics [1,2]. Furthermore, subdividing a large system differs from the original small-system thermodynamics that is based on combining separate small systems to form a large system [12–14], but also differs from the standard cellular method of subdividing a large system into cells with interactions between the cells [56]. Often, intercellular interactions are added to allow divergent correlation lengths near critical points. Thus, these intercellular interactions are added to bypass a basic constraint from assuming constant-volume cells, which can be avoided by using variable-volume regions in the nanocanonical ensemble. Our models for uncorrelated small subsystems inside macroscopic samples match the original theory of small-system thermodynamics, and mimic the measured primary response in most materials [36–49].

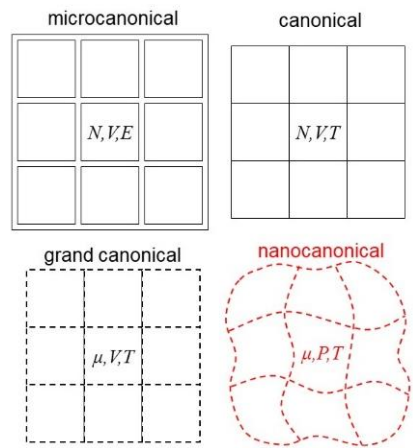

**Figure 1.** Sketch showing four ways of subdividing a sample into $\eta$ = 9 subsystems, forming various ensembles [17]. For small and fluctuating subsystems, full accuracy requires that the correct ensemble be used for the specific constraints, which may depend on the type and time scale of the dynamics.

In Fig. 1, the microcanonical ($N,V,E$) ensemble comes from assuming fully-closed subsystems, separated by walls that are impermeable, rigid, and insulating, fully isolating every subsystem from its environment, thereby conserving the number of particles, volume and energy of each subsystem. The canonical ($N,V,T$) ensemble comes from assuming subsystems separated by walls that are impermeable and solid, but thermally conducting, allowing heat to pass freely in and out, so that energy fluctuates and $T$ replaces $E$ as an environmental variable. The grand canonical ($\mu,V,T$) ensemble comes from assuming subsystems separated by solid diathermal walls that are permeable, allowing particles to exchange freely between regions, so that the number of particles fluctuates and $\mu$ replaces $N$ as an environmental variable. The nanocanonical ($\mu,P,T$) ensemble comes from fully-open *regions*, separated by "interfaces" that are diathermal, permeable, and flexible, so that volume can change as particles and heat pass in and out, allowing density to be optimized. In this nanocanonical ensemble, spontaneous changes in $\eta$ occur except at the extremum where total energy is minimized, $\mathcal{E} = (\partial E_t/\partial \eta)_{S_t,V_t,N_t} = 0$. This $\mathcal{E} = 0$ condition also gives the equilibrium distribution and average size of the regions [57]. In general, $\mathcal{E} \rightarrow 0$ for infinite systems and $\mathcal{E} = 0$ at equilibrium in the nanocanonical ensemble, but $\mathcal{E}$ is usually not zero in-between.

Various physical mechanisms could cause abrupt interfaces between regions. For semi-classical systems, such as an ideal gas, interfaces could correspond to where wavefunction decoherence breaks the quantum-exchange symmetry, allowing atoms in separate regions to be distinguishable by their distinct locations. Similarly, because ferromagnetic interactions require quantum exchange, realism in the standard Ising model (where each spin is assumed to be localized to a single site) can be improved by allowing quantum exchange between spins that are delocalized over a region. Because quantum coherence is unlikely to occur across an entire ferromagnetic sample, abrupt decoherence of local wavefunctions could again define an interface between regions. Such abrupt decoherence may be facilitated by time-averaging: fluctuations within different regions occur at different rates, so that mutual interactions across interfaces are soon averaged to zero. In any case, several experimental techniques have established that such dynamical heterogeneity dominates the primary response of most materials [36–49], consistent with nanoscale regions having relaxation rates that can differ by orders of magnitude across abrupt interfaces [58–60]. Another way to form regions may involve fluctuations in particle density due to anharmonic interactions, consistent with MD simulations [50]. In any case, the nanocanonical ensemble removes all external constraints from inside the system, allowing bulk samples to find their thermal equilibrium distribution of internal regions.

One explanation for why Hill's crucial contribution to conservation of energy has escaped broad attention is that its influence can be subtle. For example, despite the conceptual similarities between $\mathcal{E}d\eta$ and $\mu dN$, the magnitude of $\mathcal{E}$ is often less than the magnitude of $\mu$. However, only *changes* in energy are relevant to Eq. (2). For example, systems with a fixed number of particles have $\mu dN = 0$. Furthermore, one of the most important uses of chemical potential is when $\mu = 0$, yielding the thermal equilibrium of systems that have no external constraints on $N$, such as for phonon or photon statistics. Similarly, perhaps the most important use of the subdivision potential is when $\mathcal{E} = 0$, used for the nanocanonical ensemble and thermal equilibrium of systems that have no external constraints on their internal heterogeneity [57]. In the theory of an ideal gas (subsection 5.1), $\mathcal{E}$ contributes less than 6% to the total entropy per particle, whereas $\mathcal{E}$ controls 100% of the internal heterogeneity that is prohibited in standard thermodynamics. Indeed, the importance of $\mathcal{E}$ comes not from its magnitude, but from its broad applicability to a wide range of situations. In fact, because Hill's $\mathcal{E}$ is necessary for describing thermal heterogeneity and local equilibrium inside most systems, even without external changes, $\mathcal{E}$ may have a broader impact than many other parameters in thermodynamics.

Nanothermodynamics was first applied to macroscopic systems in a mean-field model of glass-forming liquids [15]. The model provides a unified picture for stretched-exponential relaxation and super-Arrhenius activation in terms of a phase transition that is broadened by finite-size effects. Another early application also utilized mean-field theory in the nanocanonical ensemble to explain non-classical critical scaling measured in ferromagnetic materials and critical fluids [16]. This mean-field cluster model treats Ising-like spins with mean-field energies that are heterogeneously localized within nanoscale regions, unlike the usual assumption of homogeneous interactions throughout macroscopic samples. Local mean-field energies may come from interactions that are time-averaged within each region, or location-averaged across the region, attributable to exchange interactions that occur only between particles within each region. The model gives excellent agreement with many measurements, including at temperatures just above the critical point where measured critical exponents increase as $T$ is increased [17,61], opposite to the monotonic decrease predicted by homogenous theories and simulations [62]. The mean-field cluster model matches the measured $T$ dependence of effective scaling exponents using $\mu/kT$ as a basic constant for each system, instead of the non-classical scaling exponent as an empirical parameter. Here we utilize the 1-D Ising model to obtain analytic expressions for idealized thermal behavior, without adjustable parameters.

## 4. Extending Statistical Mechanics to Treat Multiscale Heterogeneity

Standard statistical mechanics usually starts by calculating a partition function from a simplified model of a physical system. These partition functions are obtained by Legendre transforms [26,63] that involve summing (or integrating) over all possible states of the model, weighted by the probability of each state. An example is the canonical-ensemble partition function

$$Q_{N,V,T} = \sum_E \Omega_{N,V,E} e^{-E/kT}. \tag{5}$$

Equation (5) is used to calculate the Helmholtz free energy $A = -kT \ln Q$, the chemical potential $\mu = (\partial A/\partial N)_{V,T}$, and the average internal energy

$$\bar{E} = \partial \ln Q \, / \partial (-1/kT). \tag{6}$$

A second Legendre transform yields the partition function for the grand-canonical ensemble

$$\Xi_{\mu,V,T} = \sum_{E,N} \Omega_{N,V,E} e^{-E/kT} e^{\mu N/kT}, \tag{7}$$

which gives the grand potential $\Phi = -kT \ln \Xi$ and average number of particles $\bar{N} = -(\partial \Phi/\partial \mu)_{V,T}$. Because nanothermodynamics includes non-extensive contributions to energy, it allows a third Legendre transform into the nanocanonical ensemble

$$\Upsilon_{\mu,P,T} = \sum_{N,VE} \Omega_{N,V,E} e^{-E/kT} e^{\mu N/kT} e^{-PV/kT}, \tag{8}$$

which yields the subdivision potential $\mathcal{E} = -kT \ln \Upsilon$. Alternatively, the subdivision potential can be found by removing all extensive contributions to the internal energy, as given by Eq. (3).

Most examples presented here are based on the 1-D Ising model for binary degrees of freedom ("spins"). The model was originally used by Ernst Ising [64] in an attempt to explain ferromagnetic phase transitions using spins with a magnetic moment, $m$. Binary states of the spins come from assuming that they are uniaxial, constrained to point either "up" ($m$ in the +z-direction) or "down" ($m$ in the $-z$-direction). Ising's model applies equally well to other systems having binary degrees of freedom, such as the interacting lattice gas of occupied or unoccupied sites, or the binary alloy of two types of atoms on a lattice. The standard 1-D model in the thermodynamic limit, solved by Ising in 1925, does not have a phase transition until $T = 0$. Onsager's tour-de-force treatment of the Ising model in 2-D was the first analytic solution of a microscopic model to show a phase transition at $T > 0$. Note, however, because Onsager's solution assumes an infinite homogenous system in a specific ensemble, where each spin is distinguishable by its location without exchange between neighboring sites, it may not apply to most real systems. Nevertheless, as the simplest microscopic model having a thermal phase transition, the Ising model remains widely studied to investigate how statistical mechanics can be used to yield thermodynamic behavior. Because the spins are fixed to a rigid lattice, $P$ and $V$ play no role in the energy, replaced by the conjugate variables of magnetic field ($B$) and total magnetic moment ($M_t = \eta M$). For the simple models presented here $B = 0$, so that only two environmental variables are needed to define the ensemble. The canonical ensemble partition function becomes

$$Q_{N,T} = \sum_E \Omega_{N,E} e^{-E/kT}. \tag{9}$$

As before, $A = -kT \ln Q$ gives the free energy, and Eq. (6) the average internal energy. Similarly, the nanocanonical partition function is

$$\Upsilon_{\mu,T} = \sum_N Q_{N,T} e^{\mu N/kT}, \tag{10}$$

with the subdivision potential $\mathcal{E} = -kT \ln \Upsilon$. Alternatively, as with Eq. (3), $\mathcal{E}$ can be obtained from the Euler equation by solving for the non-extensive contributions to the internal energy

$$E = TS + \mu N + \mathcal{E}. \tag{11}$$

In what follows we first describe nanothermodynamics for the semi-classical ideal gas of noninteracting atoms. Subsequent examples involve the Ising model for interacting uniaxial spins. We consider the usual case of spins that are distinguishable by their location, but then treat spins that are indistinguishable in nanoscale regions, attributable to particle symmetry due to an exchange interaction that is localized within each region.

## 5. Finite-Size Effects in the Thermal Properties of Simple Systems

*5.1 Semi-Classical Ideal gas*

Thermal fluctuations and other finite-size effects are often assumed to negligibly alter the average properties of large systems [65–67]. However, we now show that finite-size effects may be necessary to find the true thermal equilibrium in systems of any size. First focus on a large volume ($V$~1 m³) containing on the order of Avogadro's number of atoms ($N$~$N_A$=6.022x10²³ atoms/mole). Assume monatomic atoms at temperature $T$ with negligible interactions (ideal gas), so that the average internal energy comes only from their kinetic energy, $\bar{E}$=3$N$(½$kT$). Gibbs' paradox [18–21] is often used to argue that the entropy of such thermodynamic systems must be additive and extensive. Nanothermodynamics is based on assuming standard thermodynamics in the limit of large systems, while treating non-extensive contributions to thermal properties of small systems in a self-consistent manner. Here we review and reinterpret several results given in chapters 10 and 15 of Hill's *Thermodynamics of Small Systems* [13]. We emphasize that sub-additive entropy, a fundamental property of quantum-mechanics [23,68], often favors subdividing a large system into an ensemble of nanoscale regions, increasing the total entropy and requiring nanothermodynamics for a full analysis.

Table I gives the partition function, fundamental thermodynamic function (entropy, free energy, or subdivision potential) and other thermal quantities for an ideal gas of mass $m$ in the four ensembles of Fig. 1, similar to the tables in [26]. (Subscripts on the entropy and subdivision potential denote the ensemble.) Other symbols used in Table I include the thermal de Broglie wavelength $\Lambda = h/\sqrt{2\pi mkT}$ (where $h$ is Planck's constant), and the absolute activity $\lambda = e^{\mu/kT}$. Table I elucidates several aspects of nanothermodynamics of the ideal gas in various ensembles. The microcanonical partition function

| Table I: monatomic ideal gas ||| 
| Ensemble | Partition function | Fundamental thermodynamic function and variables |
|---|---|---|
| microcanonical $(N,V,E)$ | $\Omega_1 = V\frac{\pi}{4}\left(\frac{8m}{h^2}\right)^{3/2}\sqrt{E}$ <br> $\Omega_N \approx \frac{1}{N!}\left[V\left(\frac{4\pi mE}{3Nh^2}e\right)^{3/2}\right]^N$ | $S_{mc}/Nk = \frac{1}{N}\ln(\Omega_N) \approx \frac{3}{2} + \ln(V) + \frac{3}{2}\ln\left[\frac{4\pi mE}{3Nh^2}\right] - \frac{1}{N}\ln(N!)$ <br> $\approx \frac{5}{2} - \ln\left(\frac{N}{V}\right) + \frac{3}{2}\ln\left[\frac{4\pi mE}{3Nh^2}\right] - \frac{1}{N}\ln(\sqrt{2\pi N})$ <br> $\mathcal{E}_{mc}/kT \approx \ln(\sqrt{2\pi N})$ |
| canonical $(N,V,T)$ | $Q_1 = \int_0^\infty \Omega_1 e^{-\frac{E}{kT}}dE$ <br> $= V/\Lambda^3$ <br> $\Lambda = h/\sqrt{2\pi mkT}$ <br> $Q_N = Q_1^N/N!$ | $A/kT = -\ln(Q_N) = -N\ln(Q_1) + \ln(N!)$ <br> $\approx N\ln(N\Lambda^3/V) - N + \ln(\sqrt{2\pi N})$ <br> $\bar{E} = \frac{\partial \ln Q_N}{\partial(-1/kT)} = \frac{3}{2}NkT$ <br> $\bar{\mu}/kT \equiv -\ln(Q_{N+1}/Q_N) = \ln(\Lambda^3/V) + \ln(N+1)$ <br> $S_c/Nk = (\bar{E}-A)/NkT \approx \frac{5}{2} - \ln(N\Lambda^3/V) - \frac{1}{N}\ln(\sqrt{2\pi N})$     $\mathcal{E}_c = \mathcal{E}_{mc}$ |
| grand canonical $(\mu,V,T)$ | $\Xi = \sum_{N=0}^\infty \frac{1}{N!}\left[\frac{V}{\Lambda^3}\right]^N \lambda^N$ <br> $\lambda = e^{\mu/kT}$ | $\Phi/kT = -\ln(\Xi) = -V\lambda/\Lambda^3$ <br> $\bar{N} = -\partial\Phi/\partial\mu = V\lambda/\Lambda^3$ <br> $\lambda = e^{\bar{\mu}/kT} = \bar{N}\Lambda^3/V \;\rightarrow\; \bar{\mu}/kT = \ln(\bar{N}\Lambda^3/V)$ <br> $S_{gc}/\bar{N}k = (\bar{E}-\Phi-\mu\bar{N})/\bar{N}kT = \frac{5}{2} - \ln(\bar{N}\Lambda^3/V)$     $\mathcal{E}_{gc} = 0$ |
| nanocanonical $(\mu,P,T)$ | $\Upsilon = \int_0^\infty e^{\frac{V}{\Lambda^3}\lambda}\, e^{-\frac{PV}{kT}}\left[\frac{P}{kT}\right]dV$ | $\mathcal{E}_{nc}/kT = -\ln(\Upsilon) = \ln[1 - kT\lambda/P\Lambda^3]$ <br> $\bar{N} = \lambda\partial\ln(\Upsilon)/\partial\lambda = (kT\lambda/P\Lambda^3)/[1-kT\lambda/P\Lambda^3] \;\rightarrow\; kT\lambda/P\Lambda^3 = \bar{N}/(\bar{N}+1)$ <br> $\mathcal{E}_{nc}/kT = -\ln[\bar{N}+1]$ <br> $S_{nc}/\bar{N}k = (\bar{E}+P\bar{V}-\mu\bar{N}-\mathcal{E}_{nc})/\bar{N}kT = \frac{5}{2} - \ln(\bar{N}\Lambda^3/\bar{V}) + \frac{1}{\bar{N}}\ln(\bar{N}+1)$ |

comes from the multiplicity of microscopic states that have energy *E*. Partition functions in other ensembles come from one or more Legendre transforms to yield other sets of environmental variables. If the transform involves a continuous variable, it should be done using an integral over the variable. However, if the variable is discrete (e.g. *N*), in nanothermodynamics it is especially important to use a discrete summation, thereby maintaining accuracy down to individual atoms, which also often simplifies the math and removes Stirling's formula for the factorials. Similarly, note that the chemical potential in the canonical ensemble is calculated using a difference equation, not a derivative, so that again Stirling's formula can be avoided. Another general feature to be emphasized is that the variables shown in the "Ensemble" column are fixed by the environment (e.g. types of walls surrounding a subsystem); hence they do not fluctuate. In contrast, each conjugate variable fluctuates due to contact with the environment, so that these conjugate variables are shown as averages. Thus, as expected for small systems [14], it is essential to use the correct ensemble for determining which variables fluctuate, and by how much.

    Now focus on the entropy. Recall that the Sackur-Tetrode formula for the entropy of an ideal gas is $S_0/k = 5N/2 - N\ln[N\Lambda^3/V]$. Note that to make this entropy extensive, the partition function is divided by *N*!, which assumes that all atoms in the system are indistinguishable, usually attributed to quantum symmetry across the entire system. However, the need to use macroscopic quantum mechanics for the semi-classical ideal gas remains a topic of debate [18–21]. Table I shows that in nanothermodynamics, entropy is non-extensive due to contributions from subtracting the subdivision potential $S/k = S_0/k - \mathcal{E}/kT$ (see Eq. 4). For example, in the canonical ensemble $\mathcal{E}_c/kT \approx \ln\sqrt{2\pi N}$, which comes from Stirling's formula for *N*!. Because the Legendre transformation from *N* to *μ* is done by a discrete sum over all *N*, Stirling's formula is eliminated from the grand-canonical and nanocanonical ensembles. Instead, a novel non-extensive contribution to entropy arises in the nanocanonical ensemble from $\mathcal{E}_{nc}/kT = -\ln(\bar{N}+1)$. Because this negative subdivision potential is subtracted from *S/k*, the entropy per particle increases when the system subdivides into smaller regions. This entropy increase appears only in the nanocanonical ensemble, where the sizes of the regions are unconstrained, a feature that is unique to nanothermodynamics. Figure 2 is a cartoon sketch of how net entropy may change if a single system subdivides into subsystems: decreasing if subsystems are constrained to have fixed *V* and *N* (canonical ensemble), but increasing if subsystems have variable *V* and *N* (nanocanonical ensemble). As expected, total entropy increases if most atoms can be distinguished by their nanoscale region, even if they may soon travel to other regions to become indistinguishable with other atoms. In fact, for the semi-classical ideal gas, the fundamental requirement of sub-additive quantum entropy [23,68] is found only in the nanocanonical ensemble.

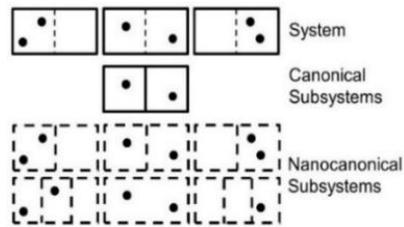

**Figure 2.** Crude characterization of a system (top row) and its multiplicities for two types of subdivision. The particles (dots) can be on either side of the system, but must be in separate volumes for fixed *N* in canonical subsystems (second row). Nanocanonical subsystems have variable *N*, and variable *V*, increasing the net entropy.

The subdivision potentials from Table I can be used to obtain the non-extensive corrections to entropy of specific atoms in various ensembles. As an example, consider one mole ($N$=6.022x10$^{23}$ atoms) of argon gas (mass $m$=6.636x10$^{-26}$ kg) at a temperature of 0 C ($T$=273.15 K), yielding the thermal de Broglie wavelength $\Lambda = h/\sqrt{2\pi mkT} \approx 16.7$ pm. At atmospheric pressure (101.325 kPa), the number density of one amagat ($N/V$ = 2.687x10$^{25}$ atoms/m$^3$) gives an average distance between atoms of $(V/N)^{1/3} = (\bar{V}/\bar{N})^{1/3} \approx 3.34$ nm, and a mean-free path of $\ell = (V/N)/(\sqrt{2}\pi d^2) \approx 59.3$ nm (using a kinetic diameter of $d$ = 0.376 nm for argon). Under these conditions the Sackur-Tetrode formula predicts a dimensionless entropy per atom of $S_0/Nk = 5/2 - \ln[\Lambda^3 \bar{N}/\bar{V}] \approx 18.39$ (equal to 152.9 J/mole-K). In the canonical ensemble the subdivision potential is positive, $\mathcal{E}_c/kT \approx \ln(\sqrt{2\pi N})$, so that when subtracted from the Sackur-Tetrode formula the entropy is reduced. Although the magnitude of this entropy reduction per atom is microscopic, $\mathcal{E}_c/NkT$ = 4.70x10$^{-23}$, even such a small reduction is used to justify the standard thermodynamic hypothesis of a single homogeneous system. However, the hypothesis breaks down if subsystems are not explicitly constrained to have a fixed size. Indeed, *regions* in the nanocanonical ensemble have a sub-additive entropy that increases upon subdivision. Specifically, $\mathcal{E}_{nc}/kT = -\ln(\bar{N}+1)$ is negative when $\bar{N} > 0$, confirming that any system of ideal gas atoms favors subdividing into an ensemble of regions whenever the size of each small region is not externally constrained. Thermal equilibrium in the nanocanonical ensemble is usually found by setting $\mathcal{E}_{nc} = 0$ [57], yielding $\bar{N} \to 0$ and an increase in entropy per atom of: $-\mathcal{E}_{nc}/\bar{N}kT = \lim_{\bar{N}\to 0}[\ln(\bar{N}+1)/\bar{N}]$ = 1, about 5.4% of the Sackur-Tetrode component. However, the Sackur-Tetrode formula has been found to agree with measured absolute entropies of four monatomic gases, with discrepancies (0.07-1.4%) that are always within two standard deviations of the measured values [69]. Thus, the experiments indicate that $\bar{N} \gg 1$ in real gases, presumably due to quantum symmetry on length scales of greater than 10 nm. For example, if quantum symmetry (indistinguishability) occurs for atoms over an average distance of the mean-free path ($\ell$ = 58.3 nm), then $\bar{N} = \ell^3(N/V) \approx 5600$ atoms. Now the subdivision potential per atom yields $-\mathcal{E}_{nc}/\bar{N}kT = \ln(\bar{N}+1)/\bar{N} \approx 0.0015$, well within experimental uncertainty. In any case, nature should always favor maximum total entropy, no matter how small the gain, so that the statistics of indistinguishable particles may apply to semi-classical ideal gases across nanometer-sized regions, but not across macroscopic volumes.

Having $\bar{N} \ll N$ for a semi-classical ideal gas implies that many atoms can be distinguished by their local region within the large system. Thus, as expected, a large system of indistinguishable atoms can increase its entropy by making many atoms distinguishable by their location. Furthermore, because the nanocanonical ensemble allows fluctuations around $\bar{V}$, local regions may adapt their size and shape to encompass atoms that are close enough to collide, or at least to have wavefunctions that may overlap, which is the usual criterion for the onset of quantum behavior. Figure 3 is a cartoon sketch depicting two ways of mixing gases from two boxes, with the color of each box representing the particle density of each type of gas. The upper-left sketch shows two boxes containing different gases, but with the same volume and particle density, that combine irreversibly with a large increase in entropy due to mixing. Whereas the lower-left sketch shows two identical boxes containing the same type of gas that combine reversibly, with negligible change in total entropy.

First consider the upper-left picture showing boxes with the same volume *V*, but different types of gases. Let one box contain $N_1$ particles of ideal gas 1, and the other box $N_2 = N_1$ particles of ideal gas 2, so that when combined, both specific densities are halved, e.g. $N_1/(V+V) = \frac{1}{2}N_1/V$. The Sackur-Tetrode formula yields an increased entropy from mixing: $\Delta S_0/k = N_1\{5/2 - \ln[\Lambda^3 N_1/(V+V)]\} + N_2\{5/2 - \ln[\Lambda^3 N_2/(V+V)]\} - N_1\{5/2 - \ln[\Lambda^3 N_1/V]\} - N_2\{5/2 - \ln[\Lambda^3 N_2/V]\} = (N_1+N_2)\ln 2$. This entropy of mixing dominates all ensembles. In fact, because the subdivision potentials in Table I depend on the number of particles in the system, but not on the volume, finite-size effects in the entropy are unchanged by mixing two types of atoms. Specifically, for the canonical ensemble: $\Delta S_c/k \approx \Delta S_0/k - 2\ln[\sqrt{2\pi N_1}] + \ln[\sqrt{2\pi N_1}] + \ln[\sqrt{2\pi N_2}] = \Delta S_0/k$. Similarly, for the nanocanonical ensemble: $\Delta S_{nc}/k = \Delta S_0/k + 2\ln(\bar{N}+1) - \ln(\bar{N}+1) - \ln(\bar{N}+1) = \Delta S_0/k$.

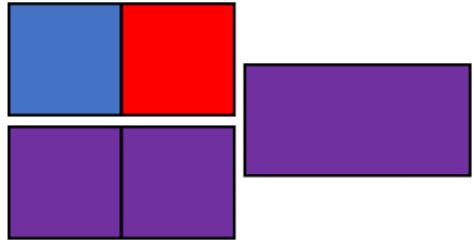

**Figure 3.** Sketch showing the process of combining two dissimilar semi-classical ideal gases (upper-left), or two similar systems of semi-classical ideal gas (lower-left). Although total particle density is constant in both cases, entropy increases due to mixing if dissimilar gases are combined.

Next consider the lower-left picture in Fig. 3 showing identical boxes, each of volume $V$ with $N_1$ particles of ideal gas 1. When the boxes are combined, the particle density does not change $(N_1 + N_1)/(V+V) = N_1/V$. From the Sackur-Tetrode formula for effectively infinite systems of indistinguishable particles, the total entropy also does not change: $\Delta S_0/k = (N_1+N_1)\{5/2 - \ln[\Lambda^3(N_1+N_1)/(V+V)]\} - 2N_1\{5/2 - \ln[\Lambda^3 N_1/V]\} = 0$. Adding finite-size effects to the canonical ensemble (row 2 in Table I), combining identical systems increases the total entropy: $\Delta S_c/k \approx \Delta S_0/k - \ln[\sqrt{2\pi(N_1+N_1)}] + 2\ln[\sqrt{2\pi N_1}] = \ln[\sqrt{\pi N_1}]$. Quantitatively, if each box initially contains one mole of particles, then $N_1 = 6.022 \times 10^{23}$ yields $\Delta S_c/k \approx 27.95$. Although the entropy increase per atom is extremely small, any increase in entropy inhibits heterogeneity in bulk systems, supporting the standard thermodynamic assumption of large homogeneous systems. However, this entropy increase applies only to ensembles having subsystems of fixed size. In contrast, combining boxes in the nanocanonical ensemble decreases the total entropy. Specifically, in thermal equilibrium at constant density, both $\bar{N}$ and $\bar{V}$ remain constant so that $\bar{N}/\bar{V} = N_1/V$, yielding a decrease in total entropy when boxes are combined: $\Delta S_{nc}/k = \Delta S_0/k + \ln(\bar{N}+1) - 2\ln(\bar{N}+1) = -\ln(\bar{N}+1)$. The per-particle entropy change is again extremely small for large boxes, but the inverse process of subdividing into small internal regions should proceed until the increase in per-particle entropy reaches its maximum: $\lim_{\bar{N}\to 0} -\Delta S_{nc}/\bar{N}k = 1$. As previously discussed (following Fig. 2), the fact that real gases do not show such deviations from the Sackur-Tetrode formula [69] implies $\bar{N} \gg 1$; but any increase in entropy is favored by the second law of thermodynamics, and required by a fundamental property of quantum mechanics for sub-additive entropy [23,68]. Moreover, similarly uncorrelated small regions are found to dominate the primary response measured in liquids and solids [36–49].

To summarize this subsection, all ensembles yield primary contributions to entropy that match the Sackur-Tetrode formula for combining ideal gases. However, nanothermodynamics allows an ideal gas to maximize its entropy and mimic measured changes in entropy, without resorting to macroscopic quantum behavior for semi-classical ideal gas atoms that may be meters apart, and therefore distinguishable by their location. Furthermore, because the nanocanonical ensemble allows the number of atoms in a particular region to fluctuate, the number of indistinguishable atoms in a specific region may be $N \gg 1$, due to atoms that are close enough to collide, or to have coherent wave functions. In any case, nature favors maximizing the total entropy whenever possible using any allowed mechanism. Hence, a novel solution to Gibbs' paradox comes from including finite-size effects in the entropy of ideal gases, without requiring quantum symmetry for macroscopic systems. This fundamental result stresses the importance of treating energy, entropy, and symmetry across multiple size scales, which requires nanothermodynamics for a fully-accurate analysis.

*5.2 Finite chain of Ising spins*

Simple models of magnetic spins provide a basic scenario for studying finite-size thermal effects between interacting particles. The fundamental equation of nanothermodynamics for reversible processes in magnetic systems is given in Fig. 4. As in Eq. (2), the equation in Fig. 4 gives changes in total internal energy of a macroscopic system from changes in total quantities, plus finite-size effects from the subdivision potential, $\mathcal{E} = (\partial E_t/\partial \eta)_{S_t,B,N_t}$. Figure 4 also shows a set of cartoon sketches of energy-level diagrams indicating how various contributions change the internal energy. Each sketch shows three energy levels, with dots depicting the relative occupation of each level. The occupation of these levels for an initial internal energy is shown by the left-most energy-level diagram. The next three energy-level diagrams, from left-to-right respectively, show that when done reversibly: adding heat ($TdS_t$) alters the relative occupation of the levels, doing magnetic work ($-M_t dB$) changes the energy of the levels, while adding spins ($\mu dN_t$) increases the occupation of all levels. The right-most energy-level diagram represents novel contributions to energy from the subdivision potential ($\mathcal{E}d\eta$). Inside a system of fixed total size ($N_t$), when the number of subsystems increases ($d\eta > 0$) the average subsystem size ($N$) decreases, the energy levels may broaden from finite-size effects due to surface states, interfaces, thermal fluctuations, etc. The subdivision potential in nanothermodynamics uniquely allows systematic treatment of these finite-size effects, thereby ensuring that energy is strictly conserved, even on the scale of nanometers.

$$dE_t = T\,dS_t - M_t dB + \mu\,dN_t + \mathcal{E}d\eta$$

**Figure 4.** Fundamental equation for conservation of energy in magnetic systems, including finite-size effects, with a sketch of how a three-energy-level system can be changed by various contributions.

The Ising model for uniaxial spins (binary degrees of freedom) demonstrates the power and utility of nanothermodynamics for finding the thermal equilibrium of finite-sized systems. Exact results can be obtained analytically in 1-D in zero magnetic field, $B = 0$, but first consider $B > 0$. Assume $N$ Ising spins, each having magnetic moment $m$ that can align in the $+B$ or $-B$ direction, with interactions only between nearest-neighbor spins. Let the spins favor ferromagnetic alignment, so that the energy of interaction (exchange energy) is $-J$ if the two neighboring spins are aligned, and $+J$ if they are anti-aligned. The usual solution to the 1-D Ising model includes contributions to energy from $mB$ and from the exchange interaction, yielding the partition function [2,3]

$$Q_{N,B,T} \approx \{e^{J/kT}\cosh(mB/kT) + [e^{-2J/kT} + e^{2J/kT}\sinh^2(mB)]^{1/2}\}^N. \tag{12}$$

If $B = 0$, the resulting free energy becomes

$$A = -kT \ln Q_{N,T} \approx -NkT \ln[2\cosh(J/kT)]. \tag{13}$$

The approximations in Eqs. (12) and (13) come from assuming large systems with negligible end effects, or equivalently spins in a ring. However, most real spin systems do not form rings, so that these equations are valid only in the limit of large systems, $N \to \infty$. We now address finite-size effects explicitly.

Consider a finite linear chain of $N+1$ spins, yielding a total of $N$ interactions ("bonds") between nearest-neighbor spins [70]. It is convenient to write the energy in terms of the binary states of each bond, $b_i=\pm 1$. Using $+J$ for the energy of anti-aligned neighboring spins, and $-J$ between aligned neighbors. The Hamiltonian is

$$E = -J\sum_{i=1}^{N} b_i, \text{ with } b_i = \{-1,+1\}. \tag{14}$$

Assuming $x$ high-energy bonds ($b_i = -1$), with ($N-x$) low-energy bonds ($b_i = +1$), the internal energy is $E = -J(N - 2x)$. The multiplicity of ways for this energy to occur is given by the binomial coefficient

$$\Omega = \frac{2N!}{x!\,(N-x)!} = 2\binom{N}{x}. \tag{15}$$

The factor of 2 in Eq. (15) is needed to accommodate both alignments of neighboring spins for each type of bond. The thermal properties of this finite-chain Ising model in various ensembles are given in Table II. Note that although the summation for the nanocanonical ensemble starts at $N = 0$, because the number of spins is $N + 1$ every region contains at least one spin, as required for spontaneous changes in the number of subsystems [57]. Also note that due to end effects, the Helmholtz free energy from Table II for $N − 1$ bonds ($N$ spins) is $A = -(N − 1)kT \ln[2 \cosh(J/kT)] − \ln 2$, equaling Eq. (13) only when $N → ∞$. Thus, if an unbroken chain is forced to have a macroscopic number of spins, all ensembles yield similar results. However, if the length of the chain can change by adding or removing spins at either end, thermal equilibrium requires the nanocanonical ensemble. As with the ideal gas, this nanocanonical ensemble is the only ensemble that does not externally constrain the sizes of the regions, so that the system itself can find its equilibrium average and distribution of sizes. From Table II, setting the subdivision potential to zero yields an average number of spins in each chain of: $\bar{N} + 1 = \cosh(J/kT) + 1$. Thus, at high temperatures the average chain contains two spins connected by one bond, whereas when $T → 0$ the average chain length diverges.

Table II: $N+1$ Ising spins ($N$ bonds) in zero field with $x$ high-energy bonds ($+J$) and ($N-x$) low-energy bonds ($-J$)

| Ensemble | Partition function | Fundamental thermodynamic function and variables |
|---|---|---|
| microcanonical ($N+1,x$) | $\Omega = \dfrac{2N!}{x!\,(N-x)!}$ | $S_{mc}/k = \ln(\Omega) = \ln[N!] + \ln(2) - \ln[x!] - \ln[(N-x)!]$ <br> $\approx N \ln(N) - x \ln(x) - (N-x)\ln(N-x) - \ln[\sqrt{2\pi x(1-x/N)}\,]$ |
| canonical ($N+1,T$) | $Q = \displaystyle\sum_{x=0}^{N} \Omega\, e^{\frac{(N-2x)J}{kT}}$ | $A/kT = -\ln(Q) = -\ln[2(e^{J/kT} + e^{-J/kT})^N]$ <br> $\qquad\quad = -N\ln[2\cosh(J/kT)] - \ln(2)$ <br> $\bar{E}/J = (2\bar{x} - N) = -N\tanh(J/kT)$ <br> $\bar{\mu}/kT = -\ln[2\cosh(J/kT)]$ <br> $S_c/k = (\bar{E} - A)/kT = -N\,(J/kT)\tanh(J/kT) + N\ln[2\cosh(J/kT)] + \ln(2)$ |
| nanocanonical ($\mu,T$) | $\Upsilon = \displaystyle\sum_{N=0}^{\infty} Q\, e^{\frac{(N+1)\mu}{kT}}$ | $\mathcal{E}_{nc}/kT = -\ln(\Upsilon)$ <br> $\qquad\quad = \ln[\tfrac{1}{2}e^{-\mu/kT} - \cosh(J/kT)] = 0 \;\rightarrow\; \mu/kT = -\ln\{2\,[\cosh(J/kT) + 1]\}$ <br> $\bar{N} + 1 = \partial \ln(\Upsilon)/\partial(\mu/kT)$ <br> $\qquad\quad = \tfrac{1}{2}e^{-\mu/kT} = \cosh(J/kT) + 1 \;\rightarrow\; \mu/kT = -\ln\{2[\bar{N} + 1]\}$ <br> $S_{nc}/k = (\bar{E} - \mu\bar{N} - \mathcal{E}_{nc})/kT = -\bar{N}\,(J/kT)\tanh(J/kT) + \bar{N}\ln\{2[\cosh(J/kT) + 1]\}$ |

As expected, Table II shows that the entropy of Ising spins increases with decreasing constraints, so that again (as in Table I) the nanocanonical ensemble has the highest total entropy. Specifically, the entropy per bond in the nanocanonical ensemble exceeds that in the canonical ensemble by the difference $\Delta S/\bar{N}k = (S_{nc} - S_c)/\bar{N}k = \ln\{[\cosh(J/kT) + 1]/\cosh(J/kT)\} - \ln(2)/\bar{N} = \ln[(\bar{N} + 1)/\bar{N}] - \ln(2)/\bar{N}$. At high $T$ where $\bar{N} \to 1$, $\Delta S/\bar{N}k \to 0$. At low $T$ where $\bar{N} \gg 1$, $\Delta S/\bar{N}k \approx 1/\bar{N} - \ln(2)/\bar{N} \to 0$. Numerical solution yields a maximum entropy difference of nearly 6% ($\Delta S/\bar{N}k$=0.0596601…) at $kT/J$ = 0.687297… where $\bar{N}$=2.25889… Hence, Ising spins in the nanocanonical ensemble always have higher entropy than if they were constrained to be in the canonical ensemble, but the excess is small at both low, and high $T$. Nevertheless, if a mechanism exists to change the length of the system, an infinite chain will shrink until there is on average $\bar{N} + 1 = \cosh(J/kT) + 1$ spins in each region, thereby maximizing the entropy of system plus its environment with no external constraints on the internal heterogeneity. In fact, because it can be difficult to fix the size of internal regions, their size should vary without external constraints, limiting the usefulness of the canonical ensemble for describing finite-size effects inside most real systems.

A key feature of the nanocanonical ensemble is that thermal equilibrium is found by setting the subdivision potential to zero [57]. Indeed, $\mathcal{E}_{nc} = 0$ ensures that the system finds its own equilibrium distribution of regions, without external constraint, similar to how $\mu = 0$ in standard statistical mechanics yields the equilibrium distribution of phonons and photons, without external constraint. Specifically, because $\mathcal{E}_{nc}$ is the change in the total energy with respect to the number of subsystems, spontaneous changes in $\eta$ occur unless $\mathcal{E}_{nc} = 0$. However, $\mathcal{E}_{nc} = -kT\ln(\Upsilon) = 0$ requires $\Upsilon = 1$ without any normalization, so that all factors must be carefully included in the partition function. For example, suppose that the factor of 2 in the numerator of $\Omega$ (Eq. 15) is ignored, from neglecting the degeneracy of each sequence of spins and its inversion. Because averages in the canonical ensemble (e.g. $\bar{E}$) are normalized by the partition function they do not change, but the average number of bonds in the nanocanonical ensemble becomes $\bar{N} = 2\cosh(J/kT)$, twice the value of $\bar{N} = \cosh(J/kT)$ from Table II.

*5.3 The Subdivided Ising Model: Ising-Like spins with a Distribution of Neutral Bonds*

Results similar to those for the finite-size Ising model in the nanocanonical ensemble (subsection *5.2*) can be obtained in the canonical ensemble by modifying the Ising model to include "neutral bonds," from nearest-neighbor spins that do not interact. (Our model differs from dilute Ising models [71] that assume empty lattice sites at fixed locations.) Physically, neutral bonds may come from neighboring spins having negligible quantum exchange (which suppresses their interaction), or from neighboring spins with uncorrelated fluctuations so that their interaction is time-averaged to zero. Again, start with the standard Ising model having $N + 1$ spins ($N$ bonds), but now let there be $\eta'$ neutral bonds (yielding $\eta = \eta' + 1$ subsystems). In addition, let there be $x$ high-energy bonds between anti-aligned spins, leaving $N - \eta' + 1 - x$ low-energy bonds between aligned spins. Figure 5 shows a specific configuration of 11 spins ($N = 10$ bonds) with $x=2$ high-energy bonds (**X**), $\eta'=3$ neutral bonds (**O**), and $N - \eta' - x = 5$ low-energy bonds (●). It is again convenient to write the energy in terms of the bonds, which may now have three distinct states, yielding the Hamiltonian

$$E = -J \sum_{i=1}^{N} b_i, \text{ with } b_i = \{-1, 0, +1\}. \tag{16}$$

The internal energy of the system is $E = -J(N - \eta' - 2x)$. The canonical ensemble involves two sums. The first sum is over $x$ for fixed $\eta'$, with a multiplicity given by the trinomial coefficient for the number of ways that the high- and low-energy bonds can be arranged among $N - \eta'$ interacting bonds. An extra factor of $2^{\eta'}$ arises because each neutral bond has two possible states for its neighboring spin. This first sum yields a type of canonical ensemble for the system with fixed $\eta'$. A second sum is over all values of $\eta'$. The multiplicity is given by the binomial for the number of ways that the neutral bonds can be distributed, which arises from the trinomial after the first summation. The behavior of this model is summarized in Table III.

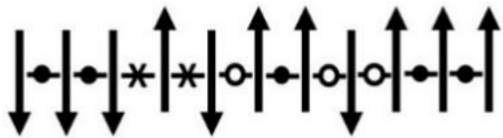

**Figure 5.** Sketch of 11 Ising-like spins in a chain connected by $N = 10$ bonds. Here $\eta' = 3$ bonds are neutral (O) (yielding $\eta = 4$ regions), $x = 2$ bonds are high-energy (X), and $N - \eta' - x = 5$ bonds are low energy (●).

| Table III: $N+1$ Ising-like spins with $\eta'$ neutral bonds (energy=0), $x$ high-energy bonds (+$J$), and $N-\eta'-x$ low-energy bonds (−$J$) | | |
|---|---|---|
| Ensemble | Partition function | Fundamental thermodynamic function and variables |
| microcanonical ($N+1, \eta', x$) | $\Omega = \dfrac{2N! \, 2^{\eta'}}{x! \eta'! (N-\eta'-x)!}$ | $S_{mc}/k = \ln(\Omega) = \ln[N!] + \ln(2^{\eta'+1}) - \ln[x!] - \ln[\eta'!] - \ln[(N-\eta'-x)!]$ <br> $\approx N \ln(N) - x \ln(x) - \eta' \ln(\eta'/2) - (N - \eta' - x) \ln(N - \eta' - x)$ |
| quasi-canonical ($N+1, \eta', T$) | $Z = \sum\limits_{x=0}^{N-\eta'} \Omega \, e^{\frac{(N-\eta'-2x)J}{kT}}$ | $Z = 2^{N+1}[\cosh(\frac{J}{kT})]^{N-\eta'} N! / [\eta'! (N - \eta')!]$ <br> $N - \eta' - 2\bar{x} = \partial \ln Z / \partial (J/kT) = (N - \eta') \tanh(J/kT)$ <br> $E/J = -(N - \eta' - 2\bar{x}) = -(N - \eta') \tanh(J/kT)$ |
| canonical ($N+1, T$) | $Q = \sum\limits_{\eta'=0}^{N} \dfrac{N!}{\eta'! \, (N - \eta')!} Z$ | $A/kT = -\ln(Q) = -(N+1)\ln(2) - N\ln[1 + \cosh(J/kT)]$ <br> $N - \bar{\eta'} = \cosh(J/kT) \, \partial \ln Q / \partial \cosh(J/kT) = N \cosh(J/kT)/[1 + \cosh(J/kT)]$ <br> $\bar{E}/J = -(N - \bar{\eta'}) \tanh(J/kT) = -N \sinh(J/kT)/[1 + \cosh(J/kT)]$ <br> $S_c/k = -N(J/kT)\sinh(J/kT)/[1 + \cosh(J/kT)] + N \ln[1 + \cosh(J/kT)]$ |

We now compare the results in Table III for the subdivided Ising model with those from Table II for the finite-size Ising model. In the canonical ensemble, the average energy of the subdivided system is higher (not as negative) as that of the finite system, as expected when neutral bonds replace an equilibrium mixture of predominantly low-energy bonds. However, the average energy per interacting bond ($\bar{E}$ from Table III, divided by $N - \bar{\eta'}$) is $-J \tanh(J/kT)$, the same as $\bar{E}/N$ from Table II. Another similarity comes from using the average number of subsystems, $\bar{\eta'} = N/[1 + \cosh(J/kT)]$, to obtain the average number of spins in each region,

$$\bar{n} = \frac{N+1}{\overline{\eta\prime}+1} = \frac{1+\cosh(J/kT)}{1+[\cosh(J/kT)]/(N+1)}. \tag{17}$$

Hence, in the limit of $N \to \infty$, Eq. (17) gives $\bar{n} - 1 \approx \cosh(J/kT)$ for the average number of bonds in the subdivided Ising model in the canonical ensemble, approaching $\overline{N} = \cosh(J/kT)$ from Table II for the finite-size Ising model in the nanocanonical ensemble. In other words, if the initial system is large enough, both approaches to nanothermodynamics are equivalent: a large ensemble of small systems (Table II) and a large system that is repeatedly subdivided into independent subsystems (Table III). Furthermore, models with distinct Hamiltonians – Eq. (16) here for a system of 3-state bonds and Eq. (14) for a system of 2-state bonds – may yield equivalent results. However, equivalence requires that the correct ensemble is used for each system, canonical ensemble here for the subdivided Ising model, and nanocanonical in subsection *5.2*. Thus, the choice of ensemble is crucial for obtaining fully-accurate behavior, even for systems that are in thermal equilibrium and in the thermodynamic limit.

*5.4 Finite Chains of Effectively Indistinguishable Ising-Like Spins*

We now consider an Ising-like model with local symmetry from particle exchange that allows nanoscale regions in the system to mimic the statistics of indistinguishable particles. Although the standard Ising model assumes that each spin can be distinguished by its location, the main mechanism causing ferromagnetism is the exchange interaction that occurs only between indistinguishable particles. Finite-size effects on the scale of nanometers inside bulk samples can be attributed to symmetry from particle exchange that is spatially localized within regions, causing the heterogeneity that defines the regions. Another consequence of particle exchange is that interaction energies are not static, i.e. the alignment between neighboring spins is not fixed, which decreases the multiplicity of degenerate energy states that become indistinguishable. Here we discuss some consequences of nanoscale symmetry on simple systems.

We again start with the finite-size Ising model comprised of a linear chain of $N$ bonds between $N+1$ spins. The Hamiltonian (Eq. (14)) yields the interaction energy $E = -J(N - 2x)$ for $x$ high-energy bonds and $(N-x)$ low-energy bonds. The average energy in the canonical ensemble is given by

$$\bar{E} = -J \frac{\sum_{x=0}^{N} \Omega(N-2x) e^{(N-2x)J/kT}}{\sum_{x=0}^{N} \Omega e^{(N-2x)J/kT}} \qquad \Omega = \begin{cases} 2\binom{N}{x} & \text{distinguishable} & (18a) \\ 2 & \text{indistinguishable} & (18b) \end{cases}$$

The binomial coefficient for the multiplicity of distinguishable states, Eq. (18a), yields the usual expression for the average energy of the Ising model, $\bar{E} = -NJ\tanh(J/kT)$, as given in Table II. However, if excited states are distinguishable by their macrostate (number of high-energy bonds $x$), but not by their microstate (location of each bond), then most states have their degeneracy reduced. We assume that all energy states are doubly degenerate, Eq. (18b), attributable to the Pauli exclusion principle for spin-½ particles, yielding results that are summarized in Table IV. Note that the average energy mimics the Brillouin function, consistent with regions having degenerate discrete states.

| Table IV: $N+1$ Ising-like spins, $N$ indistinguishable bonds: $x$ high-energy bonds (+$J$) and ($N–x$) low-energy bonds (–$J$) | | |
|---|---|---|
| Ensemble | Partition function | Fundamental thermodynamic function and variables |
| microcanonical ($N,x$) | $\Omega = 2$ | $S_{mc}/k = \ln(\Omega) = \ln(2)$ |
| canonical ($N,T$) | $Q = \sum_{x=0}^{N} \Omega\, e^{\frac{(N-2x)J}{kT}}$ | $A/kT = -\ln(Q) = -\ln\{2\sinh[(N+1)J/kT]/\sinh(J/kT)\}$ <br> $\bar{E}/J = (2\bar{x} - N) = -(N+1)\coth[(N+1)J/kT] + \coth(J/kT)$ <br> $\bar{\mu}/kT = \ln[\sinh(NJ/kT)] - \ln\{\sinh[(N+1)J/kT]\}$ $\qquad S_c/k = (\bar{E} - A)/kT$ |
| nanocanonical ($\mu,T$) | $\Upsilon = \sum_{N=0}^{\infty} Q\, e^{\frac{(N+1)\mu}{kT}}$ | $\varepsilon_{nc}/kT = -\ln(\Upsilon) = \ln[\cosh(\mu/kT) - \cosh(J/kT)] = 0$ <br> $\mu/kT = -\operatorname{acosh}[\cosh(J/kT) + 1]$ <br> $\overline{N} + 1 = \sqrt{[\cosh(J/kT) + 1]^2 - 1}$ $\qquad S_{nc}/k = (\bar{E} - \mu\overline{N} - \varepsilon_{nc})/kT$ |

*5.5 Entropy and Heat in an Ideal 1-D Polymer*

Although adiabatic demagnetization provides a well-known connection between entropy and heat in spin systems [4,41,72], this connection is often more familiar in the context of ideal polymers. Furthermore, the basic behavior of the polymer can be experienced at home using a rubber band [73].

Here, for a simplified analysis related to the 1-D Ising model, we treat an ideal polymer comprised of freely-jointed monomers (units) in 1-D [6]. Consider a polymer of $N$ units, each of length $a$. Let one end of the 1st unit be freely jointed (free to invert) about the origin ($X = 0$), with the far end of the $N^{th}$ unit unconnected. All other units have both ends freely jointed to neighboring units. For simplicity assume all units are uniaxial (1-D), with $x$ units pointing in the $-X$-direction and $N-x$ units in the $+X$-direction. The free end of the polymer is at position $X(x) = (N-2x)\,a$. The multiplicity matches that of Eq. 18a for the standard Ising model, and Table II gives the resulting microcanonical entropy $S_{mc}/k = \ln\left[2\binom{N}{x}\right]$. The elastic restoring force from the entropy [1] is: $F = -T\,\Delta S/\Delta X$ Note that this model involves differences (not differentials) because it is comprised of discrete polymer units in 1-D. Such discrete differences circumvent Stirling's formula, improving the accuracy, especially for small systems. For an incremental shortening of the polymer $\Delta X = -2a$, using half integers to best represent the average values at each integer, the change in configurational entropy of the polymer is $\Delta S_{mc}/k = \ln\left[\frac{N!}{(x+\frac{1}{2})!(N-x-\frac{1}{2})!}\right] - \ln\left[\frac{N!}{(x-\frac{1}{2})!(N-x+\frac{1}{2})!}\right] = \ln\left[\frac{N-x+\frac{1}{2}}{x+\frac{1}{2}}\right]$. Solving for the average number of negatively aligned units as a function of $F$ gives $\bar{x} = \frac{N+\frac{1}{2}(1-e^{2aF/kT})}{1+e^{2aF/kT}}$, yielding the equilibrium endpoint of the polymer $X(\bar{x}) = (N+1)\,a\tanh(aF/kT)$. At high-temperatures $X(\bar{x}) \approx (N+1)\frac{a^2 F}{kT}$, similar to the standard expression for the ideal 1-D polymer if $N \gg 1$.

This model shows a common characteristic of polymers under tension: their average length varies inversely proportional to temperature. This decrease in $X$ with increasing $T$, opposite to the behavior of most other solids, arises from the dominance of configurational entropy in polymers. Such length contraction can be observed by heating a rubber band that holds a hanging mass, demonstrating a simple conversion of heat into work. The converse conversion of work into heat can be experienced by the increased temperature of a rubber band that is rapidly stretched while in contact with your lips. Cyclic heat-to-work conversion is shown by the heat engine made from a wheel with rubber-band spokes, where an incandescent lamp causes the wheel to rotate continuously [73]. The heat-to-work mechanism comes from increased thermal agitation around the polymer, increasing its entropy and coiling it more tightly. Similarly, the work-to-heat conversion involves energy added to the heat bath when entropy is reduced as the polymer is stretched.

The purpose of this brief digression is to emphasize how changing the configurational entropy of a polymer by changing its length alters the energy of the heat bath, thus altering the entropy of the bath. We assume that an analogous change in the entropy of the polymer during a fluctuation causes a similar exchange of entropy with the bath. In other words, we assume that the heat bath cannot discern whether changes in entropy of a polymer are due to external forces, or internal fluctuations. Next, we assume that equilibrium (reversible) fluctuations occur with no net loss in entropy, so that the second law of thermodynamics is strictly obeyed. Specifically, we make the ansatz that during equilibrium fluctuations the entropy of the polymer plus the entropy of its local heat bath ($S_L(X)$) never deviate from a maximum value:

$$S(X) + S_L(X) = S_{max} \qquad (19)$$

We now expand on the concept of local heat baths in nanothermodynamics [35,74] to show how Eq. (19) facilitates reversible fluctuations. In standard thermodynamics, reversible processes must proceed at infinitesimal rates, allowing the system to couple uniformly to the effectively infinite heat reservoir. However, many thermal fluctuations are fast and heterogeneous. Nanothermodynamics is based on independent small systems, often with energy and entropy isolated from neighboring systems, consistent with the energy localization and local $T$ deduced from experiments [36–42] and simulations [50]. For the specific model presented here, imagine a large sample containing a polymer melt. Let independent polymers (or their independent monomers [49,58]) and a local heat bath occupy a nanoscale volume inside the sample. During fast fluctuations, each volume is effectively isolated from neighboring volumes, conserving local energy and local entropy (Eq. 19), characteristic of the microcanonical ensemble (upper-left boxes in Fig. 1). Note, however, that for fluctuations about equilibrium, the microcanonical boxes would have a distribution of sizes and shapes, basically a frozen snapshot of the lower-right regions in Fig. 1. During sufficiently slow fluctuations, energy and particles can transfer freely between variable volumes to maximize the total entropy and maintain a thermal equilibrium distribution of regions in the nanocanonical ensemble. Thus, accurate evaluation of thermal fluctuations often involves two ensembles, the fully-closed ensemble for fast fluctuations, and the fully-open ensemble for slow fluctuations. Partially-open ensembles (e.g. canonical and grand-canonical), which restrict the exchange of some quantities but not others, are often artificially constrained. Specifically, because excess energy is persistently localized during the primary response in liquids, glasses, polymers, and crystals [36–42], sometimes for seconds or longer, particle exchange will usually accompany these slow changes in energy. Therefore, the relatively fast transfer of energy needed for a well-defined local temperature, without also changing size and particle number, is unlikely for the primary fluctuations inside most realistic systems.

We now evaluate equilibrium fluctuations that include energy from configurational entropy, which is often ignored in standard fluctuation theory. Let there be no external force on the polymer, so that the average position of its endpoint is $X(\bar{x}) = 0$, corresponding to its maximum configurational entropy. From Eq. (19), as $S(X)$ decreases when $X \neq 0$, $S_L(X)$ must increase, and vice versa. Quantitatively, using the microcanonical entropy for the polymer (adapted from Table II), allowing fast fluctuations that are localized and reversible, the entropy of the local bath is:

$$S_L(X) = S_{max} - \ln(N!) + \ln\{[(N-X)/2]!\} + \ln\{[(N+X)/2]!\}. \tag{20}$$

We assume that Boltzmann's factor, commonly used to weight large-reservoir states, also weights the local-bath states $w_L = e^{S_L(X)/k}$. Thus, when thermally averaged, every length of the polymer is equally likely, $w_L \Omega = e^{S_L(X)/k} e^{S(X)/k} = e^{S_{max}/k}$. In other words, maintaining maximum entropy during equilibrium fluctuations removes degeneracies from systems of classical particles that have the same macrostate (e.g. same $X$, $N$, or $E$), mimicking the statistics of indistinguishable particles. In previous work it has been shown that removing the alignment degeneracy from systems with the same $X$ yields $1/f$-like noise, with several features that match the low-frequency fluctuations measured in metal films and tunnel junctions [17,74–76]. Here we describe how removing the energy degeneracy from systems with the same $E$ yields similar $1/f$-like noise at lower frequencies, combined with Johnson-Nyquist-like (white) noise at higher frequencies.

*5.6 Simulations of Finite Chains of Effectively Indistinguishable Ising-Like Spins*

We explore consequences of including contributions from configurational entropy in the total energy during equilibrium fluctuations. The manner in which we add this entropy reduces the degeneracy of most energy states, mimicking the statistics of indistinguishable particles, Eq. (18b). We study the 1-D Ising model using Monte-Carlo (MC) simulations for the dynamics. The 1-D Ising model is used for simplicity, having its multiplicity of energy states given exactly by the binomial coefficient in Eq. (18a). Although MC simulations are too simplistic for microscopic dynamics, they can accurately simulate slow thermal processes around equilibrium [77,78]. A novel ingredient in our simulations is to introduce a type of orthogonal dynamics, where changes in energy are independent of changes in alignment. Specifically, each MC step conserves energy, or alignment, with no step allowing both to change simultaneously. Such constraints on the dynamics can be justified by the fact that energy and alignment contribute to distinct thermodynamic variables, and each is governed by a separate conservation law. Analogous decoupling of degrees of freedom has been found in supercooled fluids [79,80]. This orthogonal Ising model yields a combination of $1/f$-like noise at low frequencies, and white noise at higher frequencies, similar to behavior often found in nature.

We start with a finite chain of Ising spins (subsection *5.2*), with ferromagnetic interaction $J$ between nearest-neighbor spins. The Hamiltonian is given by Eq. (14). Consider a state containing $x$ high-energy bonds and $N$–$x$ low-energy bonds. The interaction energy of this state is $E = -J(N - 2x)$, and its multiplicity is the binomial coefficient $\Omega = 2\binom{N}{x}$. Equilibrium behavior of this model in various thermodynamic ensembles is given in Table II, but these results restrict Boltzmann's factor to include only the internal energy from interactions. We now explore how adding the energy from configurational entropy alters the behavior.

The Metropolis algorithm is often used to efficiently yield the Boltzmann distribution of energy states in MC simulations. A standard MC simulation of the Ising model involves choosing a spin at random, then inverting the spin if its change in interaction energy ($\Delta E$) meets the Metropolis criterion

$$e^{-\Delta E/kT} > [0,1) \tag{21}$$

where [0,1) is a random number uniformly distributed between 0 and 1. This criterion comes from energy transfer with an ideal (effectively infinite and homogeneous) heat reservoir due to changes in interaction energy; but crucial sources of energy from configurational entropy and the local thermal bath are neglected. We therefore add a second criterion that must also be met if configuration changes

$$e^{-\Delta S(X)/k} > [0,1), \tag{22}$$

where $\Delta S(X) = S_L(X)$. Note that Eq. (22) favors high entropy in the local bath, just as Eq. (21) favors high energy (and hence high entropy) in a large reservoir. Also note that Eq. (19) gives $S_L(X) = S_{max} - S(X)$, the offset between the maximum configurational entropy and its current value, not just the change in entropy between initial and final states. Justification comes from the assumption that fast fluctuations involve local properties that do not have time to couple to the large reservoir, and even localized thermal process must not diminish the total entropy. Furthermore, $S(X)$ for finite systems involves nonlinear terms that cannot be reduced to linear differentials.

Symbols in Fig. 6 show histograms of the energy from the Ising model at four temperatures, using the standard Metropolis algorithm (open), and when contributions from configurational entropy are added (solid). Solid curves come from the integrand in the numerator of Eq. (18a), showing that using Eq. (21) alone yields the statistics of distinguishable states. Whereas dashed lines come from the integrand in the numerator of Eq. (18b), showing that adding Eq. (22) yields behavior characteristic of the statistics of indistinguishable states.

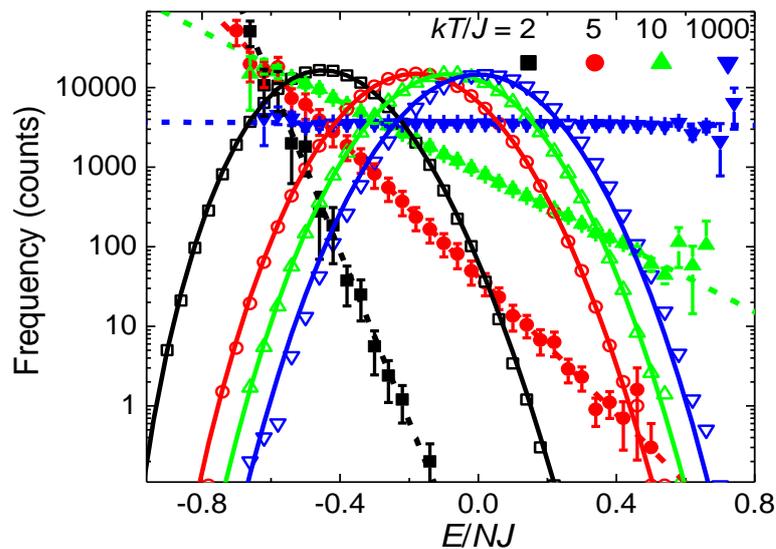

**Figure 6.** Internal energy distributions from a 1-D Ising model with N=50 bonds at four temperatures, given in the legend. Symbols show histograms from simulations using the usual Metropolis algorithm that includes only interaction energy (open), and when an offset entropy term is added (solid). Lines are fits to the data using the integrand in the numerator of Eq. (18a) (solid) or Eq. (18b) (dashed).

We now describe a dynamical sequence for simulating the Ising model which separates energy-changing steps from alignment-changing steps, thereby separating the fundamental laws of conservation of energy and conservation of angular momentum. For this orthogonal Ising model, alignment is conserved using Kawasaki dynamics [81], where neighboring spins exchange their alignments, or equivalently they exchange their locations without changing their alignments. This exchange always conserves the net alignment, but net energy often changes. Alternatively, alignment is changed without changing energy by inverting only spins that have oppositely oriented neighbors. The simulation proceeds by first choosing a spin at random from the chain, then randomly choosing whether to attempt a spin exchange or a spin flip. If spin flip is chosen, and only if the spin's neighbors are oppositely aligned, then the spin is inverted, changing the net alignment without changing the interaction energy. If instead spin exchange is chosen, and if the configurational-entropy criterion is met (Eq. (22)), then one of its nearest-neighbor spins is chosen at random. If exchanging the alignments of these two spins also meets the interaction-energy criterion (Eq. (21)) then spin exchange occurs, always preserving net alignment, but often changing the interaction energy.

Solid lines in Fig. 7 show frequency-dependent power spectral densities ($S(f)$) from simulations of the orthogonal Ising model (solid lines) and from measured flux noise in a qubit (symbols) [82]. Simulated $S(f)$ comes from the magnitude squared of the Fourier transform of time-dependent fluctuations in alignment. Note the general feature that large chains exhibit a low-frequency $1/f$-like regime that crosses over to a white noise regime at higher frequencies. Thus, this model has a single thermodynamic variable that exhibits both types of thermal noise that are usually found together in nature. The basic mechanism involves slowly-fluctuating energy (with $1/f$-like noise due to the entropy-change constraint), which slowly modulates an envelope that limits the maximum amplitude for the fast-fluctuating alignment. The orthogonal dynamics is crucial to prevent all other intermixing between distinct thermodynamic variables.

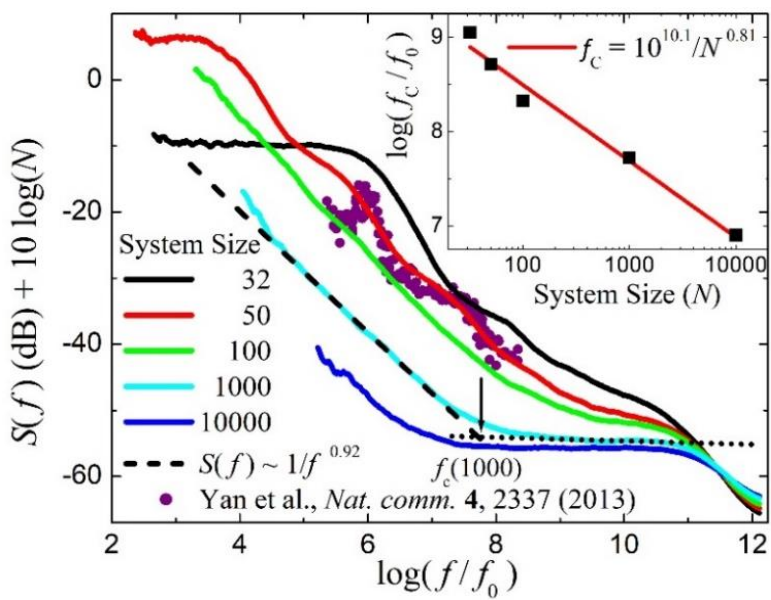

**Figure 7.** Frequency dependence of power spectral densities. Solid lines are from several simulations of an orthogonal Ising model about an average temperature of $kT/J \sim 200$, with system size ($N$) given in the legend. Solid circles show measured flux noise from a qubit [82]. The characteristic frequency ($f_0$) and amplitude of the simulations have been offset so that $N = 50$ (red line) mimics the data, with no other adjustable parameters. The dashed line fitted to the $N = 1000$ simulation has a slope of $0.92 \pm 0.02$, consistent with measurements of flux noise in qubits [83]. White noise (dotted line) occurs above the crossover frequency, $f > f_c$. The inset shows the $N$ dependence of this $f_c$.

Broken lines in Fig. 7 show linear fits to the 1000-bond chain for the $1/f$-like (dashed) and white (dotted) noise regimes. The intersection of these lines (marked by an arrow) yields the crossover frequency, $f_c(1000)$. The inset in Fig. 7 shows the chain-size dependence of this $f_c$. Three distinct features shown by the simulations in Fig. 7 mimic measured noise in quantum bits [82,83]: $1/f$-like noise with a slope of magnitude $0.92\pm0.02$; $S(f)$ in smaller chains with discrete Lorentzian spectra; and white noise at higher frequencies. Figure 7 also shows that there are two ways to reduce low-frequency noise in the orthogonal Ising model. Specifically, at $\log(f/f_0) = 4$ maximum noise occurs when $N \approx 50$. Noise decreases for larger $N$ as $f_c$ shifts to lower frequencies, and decreases for smaller $N$ as $1/f$-like noise saturates at low frequencies in small systems, avoiding the divergence as $f \to 0$ [76]. Thus, Fig. 7 shows that the orthogonal Ising model has fluctuations in alignment that yield measured frequency exponents for $1/f$-like noise, a crossover to white noise at higher $f$, and discrete Lorentzian responses; three distinct features that mimic measured spectra.

## 6. Conclusions

The theory of small-system thermodynamics is needed to ensure conservation of energy and maximum entropy in the thermal and dynamic properties of systems over multiple scales, especially on the scale of nanometers. Here we have emphasized that this "nanothermodynamics" is also crucial for obtaining the thermal equilibrium of large systems that can subdivide into an ensemble of independently fluctuating subsystems that we call "regions." These regions are fully-open – able to freely exchange energy and particles between neighboring regions without external constraints – a unique feature of the "nanocanonical" ensemble that is well-defined only in nanothermodynamics. Several results are presented to highlight various aspects of nanothermodynamics.

One result (subsection *5.1*) is to show how a large system of semi-classical ideal gas "atoms" increases its net entropy by subdividing into an ensemble of small regions. Atoms in each region are indistinguishable due to their proximity, but distinguishable from atoms in neighboring regions due to their separate locations. Thus, the need for macroscopic quantum symmetry for non-interacting point-like particles that may be meters apart is avoided. Atoms become indistinguishable only if they occupy the same nanoscale region, where they are close enough to have coherent wavefunctions, consistent with the usual criterion for the onset of quantum behavior. If two macroscopic ensembles of these regions are combined, they mimic the usual entropy of mixing for semi-classical ideal gas particles, providing a novel solution to Gibbs' paradox from finite-size effects in thermodynamics. Careful analysis reveals that the sub-additive entropy found only in the nanocanonical ensemble may be difficult to measure directly due to long-range correlations in real gases. Nevertheless, because the total entropy *decreases* when an ideal gas is subdivided into fixed volumes in the canonical ensemble, the fundamental property of quantum mechanics requiring sub-additive entropy [23,68] also favors variable volumes in the nanocanonical ensemble.

Other results presented here come from models based on Ising-like "spins," which are solved analytically in 1-D. One example (subsection *5.2*) is a chain of spins with variable length, in thermal contact with an ensemble of similar chains, which lowers its free energy by forming a nanocanonical ensemble comprised of chains with an equilibrium distribution of all possible lengths. The average length approaches two spins (one bond) at high temperatures, diverging to large systems only as the temperature goes to zero. Equivalent behavior (subsection *5.3*) is found in the canonical ensemble for a three-state model in a single chain of effectively infinite length. Subsection *5.4* describes the thermal behavior of a 1-D Ising-like model comprised of spin states with constant multiplicity, attributable to the exchange interaction between particles that makes many states indistinguishable. Subsections *5.5* and *5.6* describe how similar results are obtained by including nanoscale contributions to energy from configurational entropy during equilibrium fluctuations. Computer simulations (subsection 5.6) show that an Ising model with orthogonal dynamics (to separate changes in energy from changes in alignment) exhibits three types of thermal noise often found in nature. Specifically, the simulations show $1/f$-like noise at low frequencies, Johnson-Nyquist-like white noise at higher frequencies, with discrete Lorentzian modes visible for sufficiently small systems.

In summary, nanothermodynamics provides a systematic way to calculate contributions to energy and entropy across multiple size scales. A general result is that the correct ensemble is needed to describe the fluctuations, internal dynamics, and thermal equilibrium of macroscopic systems, regardless of system size. Furthermore, different Hamiltonians in specific ensembles can yield equivalent behavior, emphasizing the importance of including all contributions to thermal energy, not just those from microscopic interactions. Another result is a novel solution to Gibbs' paradox that has quantum symmetry of semi-classical ideal gas particles over length scales of nanometers, but not across macroscopic samples. Additional insight into the thermal behavior of other systems may come from similarly strict adherence to the laws of thermodynamics across multiple length scales.

**Acknowledgments:** We have benefited from computer simulations done by Tyler Altamirano. We acknowledge support from Arizona State University Research Computing for use of their facilities. We thank Sumiyoshi Abe for his careful reading of the manuscript, and his insightful comments.